\PassOptionsToPackage{utf8}{inputenc}
\documentclass{package/ebioinfo}
\copyrightyear{2022} 
\pubyear{2022}

\access{Advance Access Publication Date: Day Month Year}
\appnotes{Original Article}
\usepackage{amsmath}
\usepackage{multirow}
\usepackage{mathtools}
\usepackage{booktabs}
\usepackage{color, colortbl}
\usepackage{comment}
\PassOptionsToPackage{hyphens}{url}\usepackage{hyperref}
\usepackage{tablefootnote}
\usepackage{placeins}
\definecolor{gray(x11gray)}{rgb}{0.75, 0.75, 0.75}
\usepackage{multirow, makecell}
\usepackage{tabularx}
\newcolumntype{Y}{>{\centering\arraybackslash}X}
\usepackage{makecell} % for \multirowcell
\newcommand{\new}[1]{\textcolor{black}{#1}}
\newcommand{\fix}[1]{\textcolor{black}{#1}}
\newcommand{\rev}[1]{\textcolor{black}{#1}}

\begin{document}
\firstpage{1}
\subtitle{System Biology}

\title[JESTR]{\textbf{JESTR}: \underline{\textbf{J}}oint \underline{\textbf{E}}mbedding \underline{\textbf{S}}pace \underline{\textbf{T}}echnique for \underline{\textbf{R}}anking Candidate Molecules for the Annotation of Untargeted Metabolomics Data}

\author[Kalia \textit{et~al}.]{Apurva Kalia\,$^{\text{\sfb 1}}$, \fix{Yan Zhou Chen\,$^{\text{\sfb 1}}$}, Dilip Krishnan\,$^{\text{\sfb 2}}$ and Soha Hassoun\,$^{\text{\sfb 1,3,}*}$}
\address{$^{\text{\sf 1}}$Department of Computer Science, Tufts University, Medford, MA 02155, USA  \\
$^{\text{\sf 2}}$Google Research \\
$^{\text{\sf 3}}$Department of Chemical and Biological
Engineering, Tufts University, Medford, MA 02155, USA.}

\corresp{$^\ast$To whom correspondence should be addressed.}

\history{Received on XXXXX; revised on XXXXX; accepted on XXXXX}

\editor{Associate Editor: XXXXXXX}

\abstract{\textbf{Motivation:}
{A major challenge in metabolomics is annotation: assigning molecular structures to mass spectral fragmentation patterns. Despite recent advances in molecule-to-spectra and in spectra-to- molecular fingerprint prediction (FP), annotation rates remain low.
\\
\textbf{Results:} 
We introduce in this paper a novel \new{tool} (JESTR) for annotation. Unlike prior approaches that \textit{explicitly} construct molecular fingerprints or spectra, JESTR leverages the insight that molecules and their corresponding spectra are views of the same data and effectively embeds their representations in a joint space. Candidate structures are ranked based on  cosine similarity between the embeddings of query spectrum and each candidate. 
We evaluate JESTR against mol-to-spec, spec-to-FP, \new{and spec-mol matching} annotation tools on \new{four} datasets. On average, for rank@[1-20], JESTR
outperforms other tools by 55.5\% - 302.6\%. We further demonstrate the strong value  of regularization with  candidate molecules during training, boosting rank@1 performance by 5.72\% across all datasets and enhancing the model's ability to discern between target and candidate molecules. \rev {When comparing JESTR's performance against that of publicly available pretrained models of SIRIUS and CFM-ID on appropriate subsets of MassSpecGym dataset, JESTR outperforms these tools by 31\% and 238\%, respectively. }
Through JESTR, we offer a novel promising avenue towards accurate annotation, therefore unlocking valuable insights into the metabolome.
\\
}
\textbf{Availability:} Code and dataset available at https://github.com/HassounLab/JESTR1/\\
\textbf{Contact:} \href{soha.hassoun@tufts.edu}{soha.hassoun@tufts.edu}\\
\textbf{Supplementary information:} Available at \textit{Bioinformatics} online}
\maketitle 
\section{Introduction}\label{intro}
%Metabolomics is an impactful field of “omics” that involves the measurement and analysis of \textit{masses} of small molecules in biological samples \citep{wishart2016emerging}. Unlike larger molecular structures such as proteins, DNA, and RNA, small molecules have relatively low molecular weight, typically less than 1000 Da (Daltons). Small molecules are often products of metabolism, and hence referred to as metabolites. Importantly, small molecules play crucial roles in biological processes, serving as building blocks, intermediates, and regulators, and they can contribute to biomarker discovery, drug development, plant biology, nutrition, environmental health, and many applications. 

Analyzing biological samples using \textit{untargeted metabolomics}, where masses of thousands of metabolites within a biological sample are detected, presents unprecedented opportunities to characterize the metabolome. Annotation, the process of assigning chemical structures to metabolomics measurements, however is riddled with uncertainty.  Naïvely, one can presume that the measured mass could be used to determine a metabolite’s molecular structure. However, a particular molecular mass can map to possibly thousands of candidate molecular structures that share the same chemical formula. 
\new {For example, there are  44,374 known molecular structures in PubChem that are associated with chemical formula  $C_{12}H_{18}N_2O_2$)}. 
\new{Advanced computational methods promise solutions to address the challenges of annotation uncertainty.  IPA \citep{del2019integrated} and ipaPy2 \citep{del2023ipapy2} enhance confidence in metabolite identification by integrating additional data such as isotope patterns and adduct formation, to provide  statistically rigorous probability estimates for each annotation. Mummichog \citep{ li2013predicting} shifts the focus from individual metabolites to entire metabolic pathways, leveraging peak co-occurrence to infer biological relevance even when exact structures remain unknown. PUMA \citep{hosseini2020pathway}  further advances annotation by applying probabilistic modeling, predicting pathway activity and assigning chemical identities based on generative inference. Together, these tools improve annotation accuracy, enhance biological interpretation, and enable a more comprehensive understanding of the metabolome. Despite these advances based on analyzing the ionized metabolites, annotation rates remain low.}

\begin{figure*}[htbp]
\center
	\includegraphics[width=\linewidth]{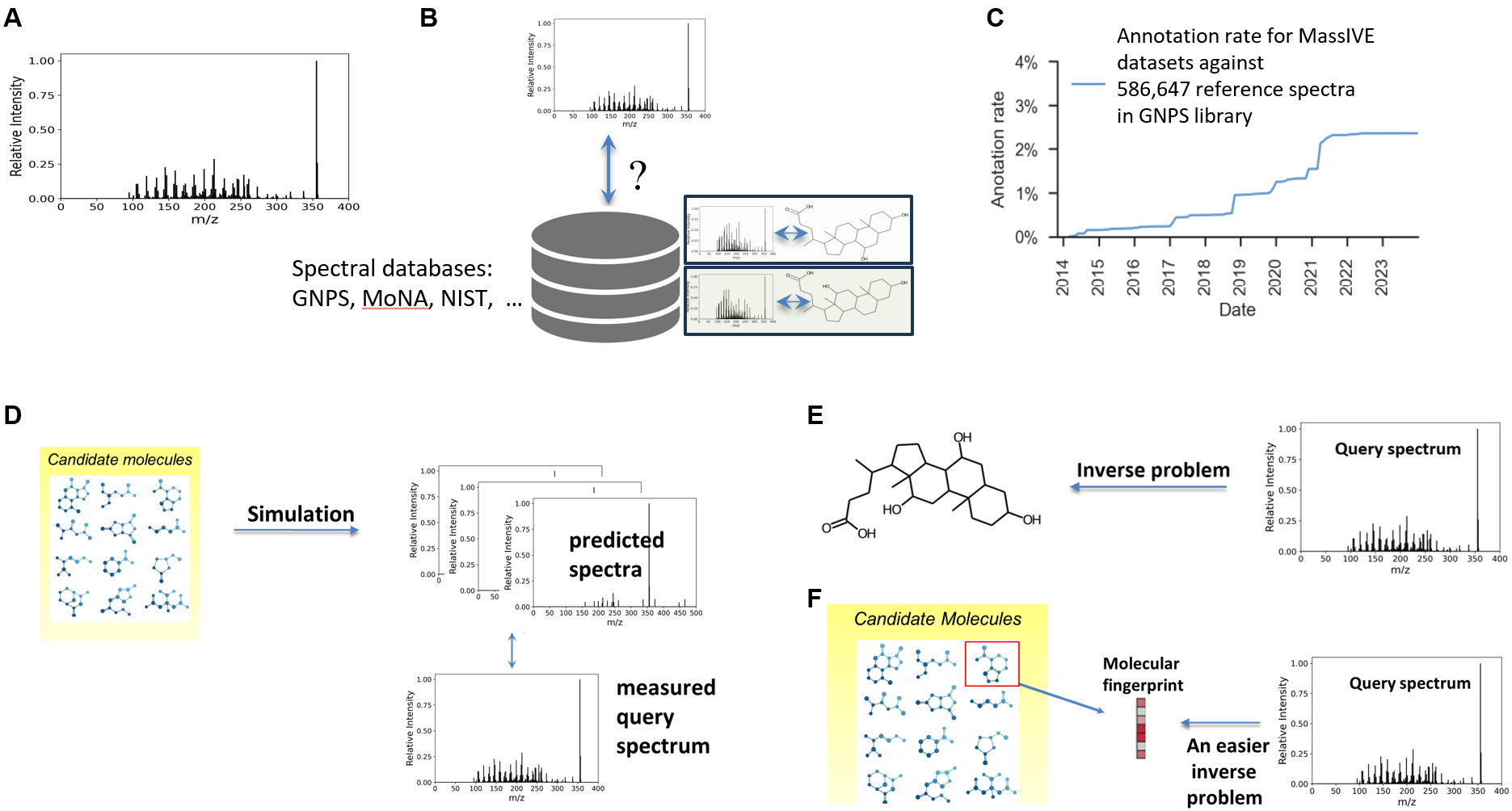}
	\caption{Current annotation workflows.   A. Example spectrum measured using LC-MS or MS/MS, where x-axis represent the mass-to-charge ratio and the y-axis represent the relative intensity of each peak. 
		B. Reference library search using spec-to-spec comparison.  
		C. Current annotation rates using state-of-the-art library search are low despite growth in reference databases.
		D. Mol-to-spec predictive approach mimics the mass spectrometry fragmentation process. 
		E. Spec-to-mol involves \textit{de novo} molecular generation from spectra and forms an inverse problem.  
		F. Spec-to-FP  approach  predicts a molecular fingerprint and identifies the candidate structure that most likely matches the predicted fingerprint.}
	\label{fig:intro}
\end{figure*}

With advances in  instrumentation, it is now possible to not only measure the mass of ionized molecules, but to also measure  masses of ionized molecular fragments. Combining liquid chromatography (LC) with mass spectrometry (MS) or combining two such analysis steps (tandem MS/MS) have now become dominant in metabolomics. The measured mass spectrum is a collection of peaks (Figure \ref{fig:intro}A). Each peak is represented by its mass-to-charge (m/z) ratio, where the charge is known and is often +1 or -1, and a relative intensity. Even for an experienced analytical chemist, assigning a chemical structure to LC-MS or MS/MS spectra is an unsolved problem as the spectrum provides a partial view on the measured molecule. Indeed, only ions that are formed by the loss or a gain of a charge can be detected by mass spectrometry. 

Several techniques address the spectra annotation problem. The  go-to technique is “spec-to-spec” comparison (Figure \ref{fig:intro}B) of the measured query spectra against  spectra that are cataloged in spectral reference libraries \citep{kind2018identification}. \new{Searching spectral libraries using learned embeddings, e.g., Spec2Vec \citep {huber2021spec2vec}, MSBERT \citep{zhang2024msbert}, DreaMS \citep{bushuiev2024emergence}, can improve search performance.}
However, despite \new{embedding advances and} growth in spectral libraries, e.g., GNPS \citep{wang2016sharing}, NIST \citep{NIST20}, MoNA \citep{MoNA}, annotation rates remain extremely low due to the imitated coverage of spectral libraries in comparison to the space of all potential molecules. In addition, measured spectra vary tremendously under differing instrument settings, e.g., ionization energy, solvent type, and adduct formation (additional functional groups attached or removed from the ionized molecule). A molecule therefore may have many corresponding spectra, which further limits library coverage. A recent search of spectra within 15,327 datasets \citep{wang2016sharing} deposited in the MassIVE (Mass Spectrometry Interactive Virtual Environment) database against 586,647 reference spectra cataloged in the GNPS reference library yielded a positive identification rate of 2.3\% \citep{martin2024molecular} (Figure \ref{fig:intro}C).  

Two types of  supervised predictive annotation techniques have emerged. “Mol-to-spec” techniques (Figure \ref{fig:intro}D) utilize combinatorial fragmentation approaches, e.g., MetFrag \citep{wolf2010silico, ruttkies2016metfrag} and CFM-ID \citep{wang2021cfm}, MLPs \citep{wei2019rapid}, or GNNs \citep{zhu2020using, li2024ensemble, young2021massformer}, to translate a molecular structure into a predicted spectrum. Candidate molecular structures are retrieved by either chemical formula, if available, or molecular mass from large molecular databases such as PubChem, or more biologically relevant, smaller, databases. The candidate with the most similar spectrum to the query spectrum is ranked highest and used as the annotation. In contrast, “spec-to-mol” techniques (Figure \ref{fig:intro}E) aim to generate \textit{de novo} molecular candidates that potentially match the query spectrum, e.g., MSNovelist \citep{stravs2022msnovelist}, Spec2Mol \citep{litsa2021spec2mol}, MS2Mol \citep{butler2023ms2mol}. For example, MS2Mol uses sequence-to-sequence transformers to translate spectra into \textit{de novo} molecular structures in the form of SMILES strings. 
 Due to their current limited capabilities, \textit{de novo} generation is presently of limited use in the metabolomics community. An alternative and earlier approach is “spec-to-FP” (Figure \ref{fig:intro}F), where a molecular fingerprint (FP) vector is predicted for the query spectrum, e.g., Sirius \citep{duhrkop2019sirius}, MIST \citep{goldman2023annotating}. Here, the predicted fingerprint is compared against those of the candidate molecular structures, and the best match, via Tanimoto  or cosine similarity, is declared  the annotation result. Despite recent advances in all such techniques, annotation rates remain low as the \textit{reconstruction} of  spectrum,   fingerprint, or  molecular structure  is a difficult task.

\begin{figure*}[t]
\begin{center}
\includegraphics[width=\linewidth]{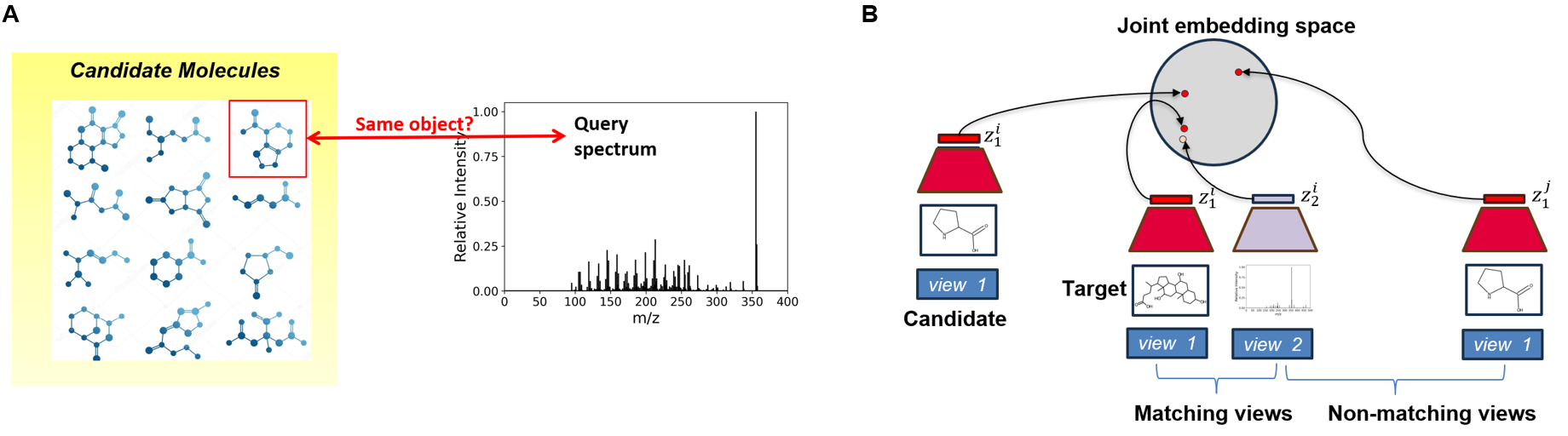}
\caption{Novelty of the JESTR annotation approach. A. JESTR avoids the explicit generation of spectra, molecules, and fingerprints, and ranks the candidate molecules against the query spectrum based on their joint-space embeddings.  B. JESTR learns to place representations of matching molecule-spectrum pairs close in the joint embedding space relative to non-matching pairs. Further, JESTR utilizes additional molecules beyond those in the training set to learn to distinguish target molecules in the training dataset from  candidate molecules (those with similar molecular formulas). \new{During ranking, the candidate whose embeddings have the maximum cosine similarity to spectrum embedding is chosen as the highest ranked molecule for that spectrum.
}}

\label{fig:fig_idea}
\end{center}
\end{figure*}

We address in this paper the problem of assigning chemical structures from a candidate set to spectral data. The novelty  in our approach lies in avoiding the explicit generation of molecular fingerprints and spectra (Figure \ref{fig:fig_idea}A),  \textit{and} in considering a molecule and its spectra as views of the same object (Figure \ref{fig:fig_idea}B). \rev { One way of avoiding the explicit construction was recently suggested by ChemEmbed \citep{faizan2025chemembed}, which  trains a spectra encoder to predict the Mol2Vec molecular embeddings  \citep{jaeger2018mol2vec}, effectively defining the embeddings  based on the Mol2Vec embedding space.
The molecule-spectra multi-modal view} insight allows us to embed molecules and their matching spectra close in the molecule-spectrum joint embedding space. Our approach avoids the need for any kind of reconstruction to intermediate forms such as fingerprints or spectra and therefore removes any reconstruction loss that would invariably creep into the ranking pipeline with any reconstruction based approach. The ranking of candidate structures can be attained by comparing their embeddings against that of the query spectrum and selecting the candidate with the highest cosine similarity. The idea of learning joint embedding spaces from multiple views of the data dates back to the seminal work on Siamese Networks \citep{chopra2005learning}.  More recently, CLIP (Contrastive Language-Image Pre-training) was trained to create a shared embedding space for both images and text, enabling the model to match relevant images and captions without the need for direct labeling or supervised training on specific dataset \citep{radford2021learning}.  As we embed  molecules and spectra in a joint embedding space,  our method is termed Joint Embedding Space Technique for Ranking candidate molecules, JESTR.
We use CMC, Contrastive Multiview Coding \citep{tian2020contrastive}, to learn view-invariant information across different views of the data and produce embeddings in a joint embedding space. 
CMC utilizes Normalized Temperature-scaled Cross-Entropy (InfoNCE) Loss, which incentivizes the model to distinguish true positive pairs from all possible negative pairs in a batch by maximizing the similarity between positive pairs while minimizing similarity with negatives. To assess the closeness of a given pair, CMC relies on a temperature-scaled cosine similarity as a discriminating function. The temperature is a model hyperparameter  that controls the sharpness of similarity scores, where lower values sharpens the distinction between positive and negative pairs.  In contrast, the recent contrastive-learning based joint-embedding model CMSSP \citep{chen2024cmssp} applies cross-entropy loss to the dot product of spectral and molecular embeddings after softmax, aiming for high embedding similarity for positives and low embedding similarity for negatives. Broadly speaking, in all contrastive learning approaches, e.g., \citep{chen2020simple, tian2020contrastive, khosla2020supervised},  the key question is how to generate the paired views (either appearing naturally or generated via data augmentation); and how to ensure that paired views end up close together in the joint embedding space. 
%Examples of pairs of input data are text/image \citep{schuhmann2022laion}, image/data-augmented image \citep{chen2020simple} and consecutive frames in a video \citep{sun2019learning}. 

%Recent work in contrastive learning \citep{chen2020simple, tian2020contrastive, khosla2020supervised} has extended the Siamese network approach significantly. 
%Non-contrastive approaches are also proving to be very effective in learning of such embedding spaces \citep{grill2020bootstrap, assran2023self, bardes2023mc, he2022masked}. 

Another novelty of our approach lies in utilizing regularization on additional data consisting of millions of molecules with the same chemical formulas  as those  in the training dataset.  While this form of data augmentation does not contribute  directly to additional labeled training data \citep{tian2022comprehensive}, the additional data is utilized to distinguish target molecules from their candidates. Here, regularization is used as a fine-tuning strategy towards the end of  training. 
When combined with contrastive loss, regularization with additional data provides two key benefits: it improves model generalization by training on a larger, more diverse set of molecules, and it enhances representation learning for both molecule and spectra embeddings by using  non-congruent pairs as additional data during training. 

We conduct experiments and analysis to answer the following research questions. (Q1) Does JESTR's implicit annotation method (with and without regularization) outperform prior explicit methods (mol-to-spec and spec-to-fp)?
(Q2) Is learning the molecule-spectrum joint embedding space effective for distinguishing  target molecules among their respective candidate sets? 
We compare JESTR against state-of-the-art mol-to-spec technique, ESP \citep{li2024ensemble},  spec-to-FP technique, MIST \citep{goldman2023annotating}, \new{and the recent mol-spec matching technique, CMSSP \citep{chen2024cmssp}}. We conduct the evaluation using \new{four} datasets: the NPLIB1 dataset that was previously released with the CANOPUS tool \citep{duhrkop2021systematic}, the well-curated, available-for-purchase NIST2020 dataset, and user-deposited data from  MassBank of North America (MoNA) \citep{MoNA}, and the recently released benchmarking datasets, MassSpecGym dataset \citep{bushuiev2024massspecgym}. \rev{
Additionally, to ensure a fair evaluation against SIRIUS and CFM-ID, we identify subsets of the MassSpecGym test set that are disjoint from their pretraining data and use these subsets to assess the performance of  JESTR relative to these tools.}

The contributions of this paper are:
\begin{itemize}
    \item Novel implicit formulation of the annotation problem to avoid  explicit prediction of spectra and fingerprints that has dominated the field since  earliest attempts in solving the problem \citep{heinonen2012metabolite}.  Our formulation is grounded in the novel insight that  molecules and spectra are views of the same object, similar to recent advances in linking text/image data. 
    \item Demonstrating that contrastive learning is effective in creating a molecular-spectra joint embedding space, and that  cosine similarity of the embeddings is sufficient for ranking  candidate molecules. That is, there is no need for an explicit (learnable) downstream ranking task.
    \item Fine-tuning the implicit model via regularization using the candidate sets of the training molecules improves the rank @1 performance in the range of \fix{ 0.51\% to 37.05\%} when compared to a baseline that does not utilize regularization. 
    \item Demonstrating that JESTR  outperforms ESP and MIST on all ranking metrics and all datasets with the exception of rank@1 for the MoNA dataset.  \fix {For rank@[1-20], JESTR outperforms ESP by 55.45\%, MIST by 56.6\%, and CMSSP by 302.56\% across four datasets}. \rev{On carefully selected subsets of the MassSpecGym dataset, JESTR outperforms SIRIUS and CFM-ID on rank@1 by 31\% and 238\%, respectively.}    These remarkable improvements are achieved even though JESTR does not utilize the additional data in the form of chemical formulae labels for spectra peaks that are currently used by MIST.
    %improves over state-of-the-art mol-to-spec technique, ESP \citep{li2013predicting}, and spec-to-FP technique, MIST \citep{goldman2023annotating}, \apurva{JESTR beats ESP performance by 37.1\%-93.2\% on all three datasets and beats MIST performance 64.2\%-83.8\% on two datasets and is behind MIST rank@1 performance by 17.6\% on the third dataset. For rank@[1 through 5], JESTR outperforms ESP by 71.6\% and MIST by 23.6\% across three datasets.}
\end{itemize}

\section{Methods}
\begin{figure}[t]
\centering
\includegraphics[scale=0.22]{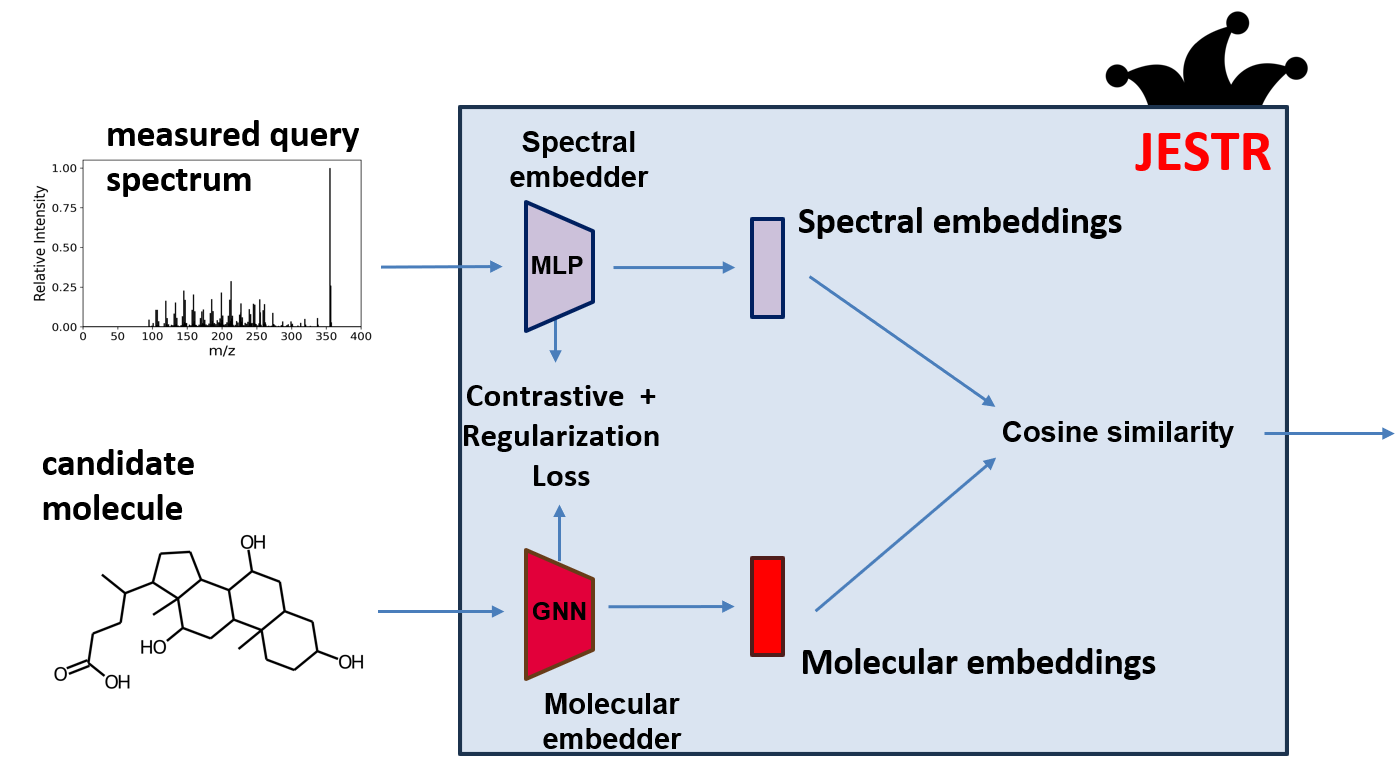}
\caption{Overview of the JESTR model architecture.  The model is trained to minimize the contrastive and regularization losses. The embeddings produced by the encoders are used to compute the cosine similarity in the joint embedding space between a molecule and a spectrum.}
\label{fig:model}
%\end{center}
\end{figure}

The JESTR model architecture (Figure \ref{fig:model}) consists of a molecular encoder and a spectral encoder. They are trained to create embeddings in a molecule-spectrum joint embedding space. To place views of the same object close to each other in the embedding space, we use the CMC contrastive learning loss \citep{tian2020contrastive}.  To improve performance, we utilize regularization. At inference, when provided a candidate set for the query spectrum, the cosine similarity is computed between each candidate and the query spectrum.  The candidates are then ranked based on their cosine similarities.  \new{Ranking results are  reported using rank@k, which is defined as the percentage of target molecules that are ranked at rank $k$  or better.
}

\subsection{Encoders}
% Molecules are provided via InChi keys. We therefore first utilizes rdkit \citep{landrum2013rdkit} to convert the InChi keys into mol structures. These 
The molecular encoder is implemented using a multi-layer Graph Neural Network (GNN) encoder. Molecular structures  are encoded as  graphs, where  node features are include atom type, atomic mass, valence, if the atom is in a ring , formal charge, radical electrons, chirality, degree, number of hydrogens and aromaticity. Edge features are the bond type, whether the bond is part of a ring, conjugacity and one hot encoding of the stereo configuration of the bond.  The encoder consists of Graph Convolutional Networks (GCNs) \citep{kipf2016semi} that aggregate information at each node. The GCNs are followed by a pooling layer and two fully connected layers to generate the final molecular embeddings,  $z_{mol}$, for a given molecule graph $c$:
\begin{equation}
    z_{mol} = MLP_{\times 2}(MAXPOOL(GCN(c)))
\end{equation}

To prepare the spectrum for its encoder, peak m/z values of the spectra are discretized into bins that are 1 Da wide. Peaks with m/z values larger than 1000 Da were dropped. The intensity are normalized to a max value of 999 - a common practice in normalizing spectral data (e.g., for the NIST datasets). For multiple peaks falling within the same bin,  peak intensities within each bin are summed  to generate the overall intensity value for that bin. A 1000-dimension binned vector therefore encodes the spectrum. A log10/3 transformation is applied to this binned vector to ensure that a few peaks and/or a long tail do not dominate the embedding vector. This 1000-dimension encoded vector was passed through a 3-layer MLP to obtain the final spectral embedding, $z_{spec}$: 
\begin{equation}
    z_s = \frac{1}{3}log_{10} ( \{\Sigma~ I_i, \forall \ i, \ n < (mz)_i < (n+1), \\
    for\ n\ in \ 0..999\}) \\
\end{equation}
where $I_i$ is the intensity of the i-th peak and ${mz}_i$ is the ${m/z}$ value of the i-th peak.
\begin{equation}
    z_{spec} = MLP_{\times 3}(z_s)
\end{equation}

\subsection{Contrastive learning of spectral and molecular views}
 We consider two views of each data item:  a molecular and a spectral view.  Each data item will have one molecular view but may have multiple spectral views as measurements may be collected under different mass spectrometry instrumentation conditions.  Matching molecule-spectrum  views  arise from a molecule and its spectrum, while non-matching views arise between a molecule and any of its non-matching spectra.   The  objective of contrastive learning on multi-views  \citep{chopra2005learning, chen2020simple, tian2020contrastive, khosla2020supervised} is to learn embeddings that separate samples from matching and non-matching distributions, and to ensure that paired views are close in the joint embedding space. 

As in CMC \citep{tian2020contrastive}, we use a discriminator function, $h$, to measure the closeness of spectral and molecular embeddings using their cosine similarity, modulated by a temperature parameter $\tau$. Thus, given a  embedding for spectrum $n$, and an embedding for molecule $m$, we can define $h$ as:
\begin{equation}
	h(z^n_{spec}, z^m_{mol}) = exp(\frac{z^n_{spec}.z^m_{mol}}{\|z^n_{spec}\|. \|z^m_{mol}\|}.\frac{1}{\tau})\label{eq:critic}
\end{equation}
The hyper-parameter   $\tau$ controls the importance of  non-matching pairs in pushing the embeddings apart in the joint embedding space.  To ensure that the discriminator assigns high values for matching pairs and low values for non-matching pairs, we define a contrastive loss, $L_{contrastive}$, over a batch of size $k$ as: 
\begin{equation}
	L_{contrastive} = \frac{1}{k} \sum\limits^k_{n=1}\left[-\mathbb{E}[log\frac{h_\theta(z^n_{spec}, z^n_{mol})}{\sum\limits^k_{m=1}h_\theta(z^n_{spec}, z^m_{mol})}]\right]\label{eq:loss}
\end{equation}
This loss effectively ensures that the cosine similarity between each matching pair,  $(z^n_{spec}, z^n_{mol})$, is highest among all possible pairings, ($z^n_{spec}, z^m_{mol}$), within the batch.

\subsection{Regularization}
As candidate molecules typically have the same molecular formula as the target molecule, we fetch such candidates from the PubChem database, and utilize regularization to train the model to better distinguish between target molecules and their candidates.  Our  regularization objective is therefore to push  candidates away from the spectra, and hence from the corresponding target, in the joint embedding space. Training using regularization is implemented by introducing an additional loss to minimize the cosine similarity between the embeddings of each spectrum and the candidate molecules of the corresponding target. Candidate sets are sorted by their Tanimoto similarity to their respective target molecule. 
For each spectrum within a training batch, we chose a set of candidates given by the batch parameter, $k_{aug}$. 
Candidates selected for regularization in each batch are therefore the $k_{aug}$ most similar candidates, and are taken sequentially in each training epoch.  Figure S1 in the Supplementary Information demonstrates the batching process for computing the regularization and contrastive losses. We then explored when and how to incorporate the regularization loss with our contrastive loss. Since our final ranking predictions are made using the molecular-spectral similarity in the joint space, the regularization attempts to push the most similar candidates away from the target molecule by minimizing a regularization loss function in addition to the contrastive loss. The regularization loss function minimizes the cosine similarity between the most similar candidates and the associated spectra - and hence the associated target molecules. The regularization loss, $L_{regularization}$, is defined as: 
\begin{equation}
    L_{regularization} =  \frac{1}{k}   \sum\limits_{n=1}^k\ \frac{1}{k_{aug}} \sum\limits_{m=1}^{k_{aug}} (cosine\_sim(z_{spec}^n, z_{cand}^m))
\end{equation}
The total training loss, $ L_{total} $, is the sum of the two losses weighted by hyper-parameters $\alpha$ and $\beta$:
\begin{equation}
	L_{total} = \alpha * L_{contrastive} + \beta * L_{regularization} \\
\end{equation}
We explored values for the $\alpha$ and $\beta$ parameters, and we observed that  utilizing regularization as a fine tuning strategy towards the end of the training provided the best performance. Regularization was turned on for the last 3\% of the training epochs. The weight given to regularization loss was 10\% to ensure that the matching pairs are not pushed too far apart during regularization. Therefore, $\alpha = 1.0$, $\beta = 0.0$\ for\ first\ 97\%\ epochs and $\alpha = 0.9$, $\beta = 0.1$\ for\ last\ 3\%\ of epochs. Implementation details and various hyperparameters are provided in Section S2 of Supplementary Information.

\section{Results}
\subsection{Datasets}\label{section:dataset}
%There are currently no standard benchmarks for evaluating the annotation problem, and prior tools utilize various curated, sometimes merged, versions of publicly, private, and for-purchase datasets. Further, some commonly used tools within the community are not available in the public domain for evaluation on common datasets, e.g., SIRIUS \citep{duhrkop2019sirius}. 
\new {Four datasets were used to evaluate JESTR.}
The NPLIB1 dataset, or the CANOPUS dataset, was first utilized by the CANOPUS tool to predict compound classes, e.g., benzenoids, phenol ethers, and others, from spectra, thus providing partial annotation on spectra in cases when spec-to-spec comparisons in reference spectral databases yield unsatisfactory matches \citep{duhrkop2021systematic}.  This dataset was created by selecting spectra from the NIST2020 \citep{NIST20}, GNPS \citep{wang2016sharing} and MoNA \citep{MoNA} databases. The selection ensured a desired distribution of compound classes. This dataset was recently renamed to NPLIB1 \citep{goldman2023annotating} to distinguish it from the CANOPUS tool. We utilized the NPLIB1  data as assembled by MIST \citep{goldman2023annotating}. This dataset comprised 8,030  spectra measured under positive mode (positively charged, with an H adduct, [M+H]+) belonging to 7,131 unique target molecules. We utilize the same data split as proposed by MIST, where the split was structure-disjoint such that a molecule with the same InChiKeys did not appear both in the training and test sets. Therefore, 714 molecules and their 819 spectra were utilized for testing. Two additional datasets were utilized to explore training JESTR on larger datasets. The NIST2020 dataset is well-curated spectral database released by the National Institute of Standards and Technology. NIST2020 comprises a variety of molecules from human, bacteria, environmental, plant, and food samples. A variety of instruments and settings are used to measure  spectra for each compound. The measurements are repeated and a consensus spectrum is created for each measurement.  The NIST datasets are available under a commercial license, and we had access to the NIST2020 version of this dataset. The MassBank of North America (MoNA) is a collaborative database, with contributions by users.
Both experimental and in-silico spectra are accepted. Here, we only retrieved the experimental dataset.  Statistics for the three datasets is provided in Table \ref{tab:datasets}. The number of unique molecules is largest in NIST2020, while the number of spectra per molecule is the lowest in NPLIB1. The splits for the NIST2020 and MoNA were created ensuring that no molecules overlapped between the training and test sets. 

\new{The fourth dataset is the recent MassSpecGym \citep{bushuiev2024massspecgym} dataset, which was developed as a benchmark set for candidate ranking, spectra simulation and molecular generation tasks. The spectra consisted of a mix of H and Na adducts, with a total of 231,104 spectra for 32,010 molecules. of high-quality labeled MS/MS spectra collected from public repositories such as MoNA, MassBank, and GNPS, as well as newly measured in-house data. The dataset includes 231,000 spectra corresponding to 29,000 unique molecular structures, making it the largest publicly available MS/MS benchmark. To ensure rigorous evaluation and prevent data leakage, the dataset is split using Maximum Common Edge Subgraph (MCES) distance, which clusters structurally similar molecules together in the same fold. Unlike InChIKey-based splits, this approach ensures that molecules in the test set have a significant structural difference from those in the training set. For the test split, an MCES distance threshold of > 10 was used. Hence, the MassSpecGym test set is more challenging than all other test splits.
}

\new{For all datasets except MassSpecGym,} we select  candidates for each target molecule in the training and test data by retrieving molecules from PubChem \citep{kim2019pubchem}, by matching formulae of the target molecules. The average candidate sets for the target molecules range from 1,322 to 2,494 molecules, and for regularization, the average candidate sets for training molecules ranged from a 1,390 to 2,322 (Table \ref{tab:datasets}). \new {For the MassSpecGym dataset, we utilize the candidate sets provided with the released benchmark. Two types of candidates are provided: either based on mass, or on formula of the precursor ion. The maximum number of candidates for each test spectrum is 256. }

\begin{table*}[th!]
\centering
\caption{Spectra, molecule, and candidate statistics for the three datasets.} 
{\begin{tabular}{@{}lcccccccccc@{}}\toprule 
    \multicolumn{1}{r}{} & \multicolumn{2}{c}{Total} 
 & \multicolumn{4}{c}{Train} & \multicolumn{4}{c}{Test}  \\
    \cmidrule(lr){2-3}\cmidrule(lr){4-7}\cmidrule(lr){8-11}
    %\midrule
         &  &   &  &  & Maximum \# & Average \# &  &  &  Maximum \# & Average \#\\
        
    Dataset & Spectra & Molecules  & Spectra & Molecules &  of candidates &  of candidates & Spectra & Molecules & of candidates   &  of candidates\\
    \midrule
    NPLIB1 & 8,030 & 7,131 &  7,211 & 6,417 & 44,374 & 2,220 & 819 & 714 & 25,929 & 2,274\\
    NIST2020 & 291,515 & 22,001 &  262,408 & 19,800 & 42,542 & 1,390 & 29,107 & 2,201 & 42,376 & 1,322\\
    MoNA & 35,752 & 6,767 &  32,216 & 6,090 & 42,542 & 2,322 & 3,536 & 677 & 32,364 & 2,494\\
    MassSpecGym & 231,104 & 32,010 & 194,119 & 28,840 & 48,184 & 2,720 & 17,556 & 3,170 & 256 & 185\\
\hline 
\end{tabular}\label{tab:datasets}}{}
\end{table*}

\begin{table*}[]
\centering
\caption{Ranking results for the NPLIB1, NIST2020,  MoNA, and the MassSpecGym datasets.  Ranking performance of ESP, MIST, CMSSP, JESTR, and  JESTR$_{\text{NR}}$,  where regularization is removed.}
\begin{tabularx}{0.95\textwidth}{cXXXXX>{\columncolor{gray}}X>{\columncolor{gray}}c}

%\begin{tabular}{c|l|cccc||c}

\hline
%\multicolumn{1}{l}{}                   & \multicolumn{1}{l}{} & \multicolumn{4}{c}{(A) Comparative results} & \multicolumn{1}{c}{(B) Ablation Study}                                \\
%\hline
&  & \multicolumn{2}{c}{Explicit-construction approaches} &\multicolumn{4}{c}{Implicit/Matching approaches}\\
 \cmidrule(lr){3-4}\cmidrule(lr){3-4}\cmidrule(lr){5-8}
Dataset& Rank & ESP  & MIST          & CMSSP (retrained) & CMSSP (pretrained)         & JESTR& JESTR$_{\text{NR} }$\\
\hline
\midrule\multirow{3}{*}{NPLIB1}        & @1           & 23.69 & 27.96       &13.92          & \textbf{54.09} & 45.76                        & 41.06                      \\
                                       & @5           & 44.69 & 62.39       & 32.97          & 67.40           & \textbf{81.53}               & 81.12                      \\
                                       & @20          & 60.20 & 82.05       & 49.08          & 74.48          & 95.77                        & \textbf{96.13}             \\
\midrule\multirow{3}{*}{NIST2020}      & @1           & 20.47 & 20.95            & 1.90         & 2.45           & \textbf{38.62}               & 36.38                      \\
                                       & @5           & 30.12 & 57.36          & 3.95       & 4.01           & \textbf{59.64}               & 56.45                      \\
                                       & @20          & 48.76 & 78.83          & 5.34       & 4.86           & \textbf{81.03}                 & 77.94                      \\
\midrule\multirow{3}{*}{MoNA}          & @1           & 19.37 & \textbf{32.30} &3.44        & 7.64           & 26.56                        & 19.38                      \\
                                       & @5           & 38.37 & 54.93          & 7.13       & 11.49          & \textbf{64.21}               & 52.82                      \\
                                       & @20          & 53.04   & 75.28          & 10.03        & 12.82          & \textbf{91.16}               & 83.88                      \\
\midrule\multirowcell{3}{MassSpecGym \\ Candidates by Mass}     
                                        & @1           & 10.71 & 14.64          &2.60       & 3.85           & 15.13                       & \textbf{15.62}             \\
                                       & @5           & 24.84 & 34.87          &5.50        & 5.44           & 36.75                       & \textbf{37.47}            \\
                                       & @20          & 42.66 & 59.15          &8.70        & 7.71           & 60.32                       & \textbf{60.55}            \\
\midrule\multirowcell{3}{ MassSpecGym \\ Candidates by formula} 
                                        & @1           & 11.05   & 9.57           &2.30     & 3.61           & \textbf{11.85}              & 11.82                     \\
                                       & @5           & 27.42 & 22.11          & 5.10       & 5.66           & 32.95                       & \textbf{33.48}            \\
                                       & @20          & 52.20 & 41.10          &8.70        & 8.53           & \textbf{61.46}              & \textbf{61.46}          \\ \hline
\end{tabularx}
\label{tab:results}%
\end{table*}

\subsection{Other annotation tools}
\new{We selected three recent annotation tools to provide a comparative evaluation for JESTR: ESP \citep{li2024ensemble},  MIST \citep{goldman2023annotating} and CMSSP \citep{chen2024cmssp}.} 
\new{ESP and MIST are recent representative methods in the mol-to-spec and spec-to-mol categories, respectively, following the approach of explicit generation of intermediate forms, while CMSSP is the only model other than JESTR that matches spectrum and candidates directly in the embedding space. 
}
The ESP model utilizes a GNN-based molecular encoder and an MLP on the molecular fingerprint. ESP is trained to learn a weighting between the molecular and fingerprint representations to predict the spectra. The best ESP performing model on rank@1 was the version that utilized the fingerprint and modeled peak co-dependencies, ESP MLP-PD. 
MIST first assigns chemical formulas to peaks within each spectrum using SIRIUS \citep{duhrkop2019sirius}, and represents a spectrum as a set of chemical formulas. 
MIST trains a transformer model to learn peak embeddings and to predicts fingerprint. MIST also  featurizes pairwise neutral losses and   predicts substructure fragments as an auxiliary task. \new{CMSSP is similar to JESTR in utilizing contrastive learning in a joint embedding space, but utilizes a discriminator function based on the dot product. }

\new{To streamline the comparison using the same data splits for all datasets, we trained  MIST and ESP on all datasets, and confirmed the results with the respective teams.  We  evaluated CMSSP on all test datasets using CMSSP's pretrained released weights. The released CMSSP was previously trained on  MS/MS spectra from two
public databases, GNPS and MassBank, and was supplemented by
1,906 spectra independently acquired in-house. To ensure that there is no data leakage between the training set and test set molecules, we also publish results by retraining CMSSP on the same training-test split as was used when reporting performance for the other models.} \new{ESP was trained for 100 epochs with a learning rate of 0.001 and batch size of 32. Adam was used as the optimizer with L2 norm of $10^{-6}$. MIST was trained for 500 epochs with a learning rate of 0.00057 and batch size of 32. For CMSSP, we evaluate the model using the released pretrained weights and also using retrained models on the specific datasets.  
}

\begin{figure*}[tbh!]
\center
	\includegraphics[width=.85\linewidth]{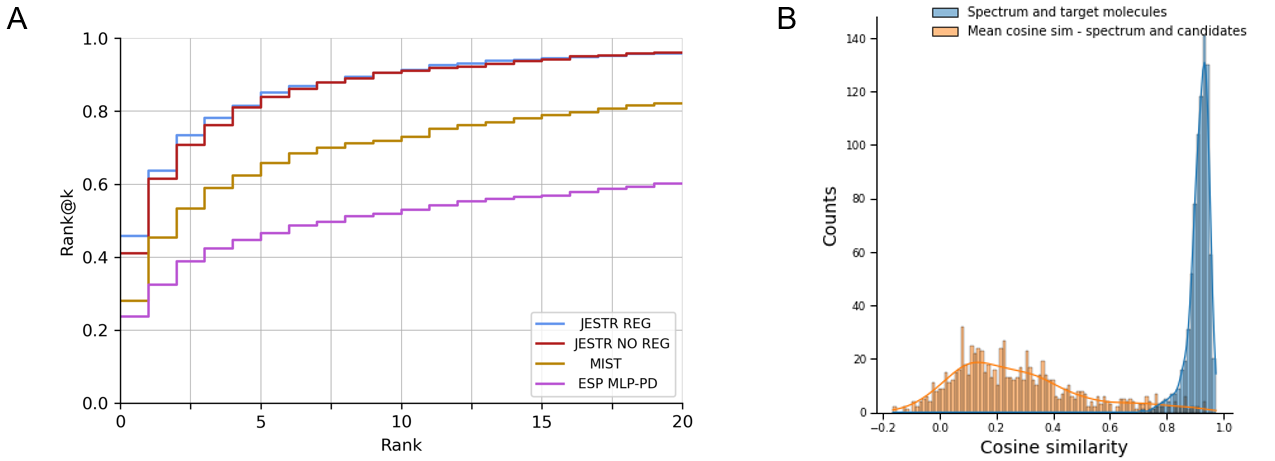}
	\caption{Results on NPLIB1. A. Rank@k results for JESTR , with and without regularization, ESP MLP-PD, and MIST. B. Distribution of cosine similarities of query spectra and target/candidate molecules  with contrastive learning using JESTR. }
\label{fig:res_explicit}
\end{figure*}
\subsection{JESTR vs explicit-construction models}
 Given a query spectrum, the primary task of JESTR  is to identify the target molecule among a set of candidates. Candidate ranking was therefore selected as  the performance    metric.  The rank@1, rank@5 and rank@20  indicates the percentage of  target molecules that were correctly ranked within the top 1, 5 and 20 candidates, respectively.  
 
 JESTR is compared against ESP and MIST (Table \ref{tab:results}). 
 For the NPLIB1 dataset, JESTR outperforms ESP and MIST on all reported ranks. For rank@1, JESTR outperforms ESP by 93.2\%, and MIST by 64.1\%. Further, JESTR achieves 95.8\% rank@20, while the maximum performance of  ESP and MIST is at 60.2\% and 82.1\%, respectively.
We  plot the detailed ranks for the NPLIB1 dataset (Figure \ref{fig:res_explicit}A). At all ranks, JESTR provides superior performance to both ESP and MIST. 
For the NIST2020 dataset, JESTR outperforms all other models. Specifically, JESTR outperforms MIST at rank@1 by 83.8\%. Detailed ranks for NIST2020 is provided in Figure S2A in the Supplementary Information. 
For the MoNA dataset, JESTR consistently outperforms ESP. MIST only outperforms JESTR at rank@1 by 17.6\%, but not on rank@5, rank@20, or any other rank, as provided in Figure S3A in the Supplementary Information. \fix{On the MassSpecGym dataset, JESTR achieves 15.62\% rank@1 using candidates by mass, a 45.9\% improvement over ESP and 6.7\% improvement over MIST. Using candidates by formula, JESTR achieves 11.85\% rank@1, surpassing ESP by 7.2\% and MIST by 23.8\%.}

Examining the overall performance of JESTR \fix{against ESP and MIST, } over the \new{four} datasets, \fix{on average for rank@[1-20], JESTR outperforms ESP  by 55.5\% and MIST by 56.6\%.}
JESTR's performance was worse on the MoNA dataset when compared to NPLIB1 and NIST2020. We suspect that JESTR's performance on  MoNA  was low for two reasons.  First,  MoNA  has the fewest number of molecules, thus providing the lowest molecular diversity among the three datasets.  Second,  MoNA data is uploaded by users and may not have undergone consistent curation efforts. The NIST2020 dataset is well curated; however, it has the highest ratio of spectra per molecule, an average of 13.25 spectra per molecule, versus 1.13 and 5.28 spectra per molecule for NPLIB1 and MoNA, respectively. As such, we suspect that JESTR finds it hard to place all the spectra embeddings closer to the molecule in the joint space for NIST2020 and MoNA. A combination of high molecule to spectra ratio, lower diversity of molecules and inconsistent spectra curation  makes JESTR perform the lowest on MoNA. MIST, with additional information in the form of  subformulae annotation on peaks, does a better job at distinguishing among various spectra of the same molecule for MoNA, thereby attaining a better rank@1 score on this dataset. However, MIST loses its advantage over JESTR starting with rank@2.

\rev {As SIRIUS and CFM-ID are widely used within the community, we benchmark JESTR's performance against their publicly available pretrained models. To ensure a fair evaluation, we focus our evaluation on the MassSpecGym benchmark dataset using candidates by formula. Among the datasets, MassSpecGym offers the most equitable basis for comparison, as the distribution of MCES distances between the pretrained molecules in SIRIUS and CFM-ID and the test molecules in MassSpecGym closely matches that of the dataset's train/test split (Figures S6-S8). 
However, because some test molecules were included in the training of SIRIUS and CFM-ID, these models may benefit from an advantage not available to JESTR. To address this, we report results on both the full test set and a filtered subset excluding any molecules seen during baseline model pretraining.}

\rev {For SIRIUS, we create a custom database containing all candidates in the test dataset. We use the GUI version 6.1.0 and ensure the correct adduct and instrument setting for each spectrum is used. Out of 17,556 test spectra, SIRIUS assigned at least one structure to 7,766 spectra, with 3,342 of those with the target molecule ranked in position 1, that is a 19.04\% rank@1 performance compared to JESTR at 11.85\%. Excluding test spectra whose molecules were in SIRIUS’s training set yields 1,281 spectra correctly annotated out of 12,314 spectra, a 10.41\% rank@1 vs 13.65\% for JESTR. This result translates to a 31.12\% improvement over SIRIUS, indicating  that JESTR generalizes better than SIRIUS on unseen molecules.}

\rev {For CFM-ID, per the authors' recommendation, we combine 3 spectra of low (<20eV), medium ($\geq$ 20eV and <40eV), and high ($\geq$ 40eV) collision energies per test molecule. Spectra are chosen randomly within the threshold, and we skip a spectrum of a specific collision energy threshold if there is not a spectrum whose collision energy falls within the threshold. Since JESTR is not designed to rank based on a set of spectra, we report the averages for best (lowest rank) and  median performance across the chosen spectra. Out of 3,170 merged spectra, CFM-ID correctly annotates 62 spectra, achieving 1.96\% rank@1 vs 14.10\% rank@1 based on the best performance and 6.56\% rank@1 based on the median rank for JESTR. Removing test spectra whose molecules were in the training set yields 60 spectra correctly annotated out of 3,004 spectra, a 2.00\% rank@1 vs 14.61\% rank@1 based on the best rank and 6.76\% rank@1 based on the median rank for JESTR. Based on the median rank, JESTR provides a 238.00\% improvement over CMF-ID.}

 The MassSpecGym benchmark dataset is notably more difficult than other datasets for all models, as molecules in the test set drastically differ from those in the training set. Nevertheless, JESTR's superior performance against explicit-construction models highlights the effectiveness of an implicit model using the joint embedding space for ranking candidate molecules.

\subsection{JESTR vs implicit models}
\rev{ Similar to JESTR, CMSSP learns a joint embedding space for spectra annotation using contrastive learning. We report CMSSP’s } performance when training on  individual datasets and using the released model weights, as training CMSSP on individual datasets shows low performance. JESTR significantly outperforms CMSSP across all datasets except for rank@1 on the NPLIB1 dataset against the pretrained CMSSP model (Table \ref{tab:results}). The strong performance of CMSSP likely results from similarities, or potentially data leakage, between our NPLIB1 test set and CMSSP training data.  

%, which typically necessitates extensive data to train well.} 

\fix{
We suspect CMSSP performed poorly when it was retrained because of its large model architecture, requiring the training of a large number of parameters. Specifically, CMSSP uses a GNN-based encoder for the molecule together with an MLP for the  molecular fingerprint while using a transformer-based encoder with 8 attention heads for the spectra. 
Two other  differences between CMSSP and JESTR is the loss function and the use of the dot product vs cosine similarity to evaluate embedding similarities. 
}
\fix{
We evaluate the impact of these differences on the NPLIB1 dataset (Table \ref{tab:CMSSP_loss}).
We first train the JESTR architecture with CMSSP's cross entropy (CE) loss and the dot product. 
Performance drops substantially when swapping the CMSSP architecture for JESTR's architecture. 
Based on performance, the CMSSP architecture is more suited for  using CE loss and the dot-product when compared to the JESTR architecture. 
Maintaining the CE loss, we then evaluate the effect of swapping the cosine ranking to replace the dot-product ranking. 
Using cosine similarity for ranking surpasses using the dot product and improves rank@5 and rank@20 performance over the CMSSP model by 22.5\% and 36.8\%, respectively, even though it was trained using the dot product. Cosine similarity explicitly normalizes the embedding magnitudes, yielding a competitive edge: similarity comparisons rely purely on embedding direction rather than magnitude. Embeddings with artificially large magnitudes are prevented from dominating similarity scores. InfoNCE with temperature-scaled cosine similarity is more suitable for a contrastive learning framework, as the normalization from cosine similarity reduces the influence of magnitude differences and the temperature scaling allows for fine-grained control over the distribution of similarities. The combination of a simple architecture and InfoNCE with temperature-scaled cosine similarity provides JESTR a superior performance compared to CMSSP.}

\begin{table*}[th!]
    \centering
    \caption{Ranking result on the NPLIB1 dataset, contrasting the CMSSP architecture against the JESTR architecture with varying losses: cross entropy (CE) and InfoNCE losses, and ranking with dot product (DP) and cosine similarity.}
 \begin{tabularx}{0.7\linewidth}{XYYYY}
    \toprule
    & \multicolumn{1}{c}{ CMSSP Architecture} &  \multicolumn{3}{c}{JESTR Architecture}\\
\cmidrule(lr){2-2} \cmidrule(lr){3-5}
& CE loss  & CE loss & CE loss&  InfoNCE loss \\
Rank (\%) & DP ranking  & DP ranking & cosine ranking & cosine ranking   \\
    \midrule
         rank@1 & 13.92  & 3.20 & 10.84  & \textbf{45.76}\\
         rank@5 & 32.97  & 16.38 & 40.39  & \textbf{81.53}\\
         rank@20 & 49.08 & 37.68 & 67.12  & \textbf{95.77}\\
    \bottomrule
    \end{tabularx}
    \label{tab:CMSSP_loss}
\end{table*}

 \subsection{Joint-space embeddings distinguish target molecules from their candidates}
 The  contrastive loss used in training JESTR ensures that the embeddings for matched spectrum-molecule pairs are placed close to each other in the joint embedding space, while non-matching spectrum-molecule pairs are placed further away.  Figure \ref{fig:res_explicit}B shows the distribution on  spectrum-molecule cosine similarities for matching and non-matching pairs in the NPLIB1 test set. The corresponding distributions for NIST2020 and MoNA datasets are shown in Figure S2B and S3B respectively in the Supplementary Information. While any molecule other than the target can be considered a non-matching partner for the query spectrum, Figure \ref{fig:res_explicit}B only considers candidate molecules (same chemical formula as the target) as the non-matching partner. It is clear that JESTR well  discriminates between target and candidate molecules.
  \begin{figure*}[bth!]
  \begin{center}    
  	\includegraphics[width=\linewidth]{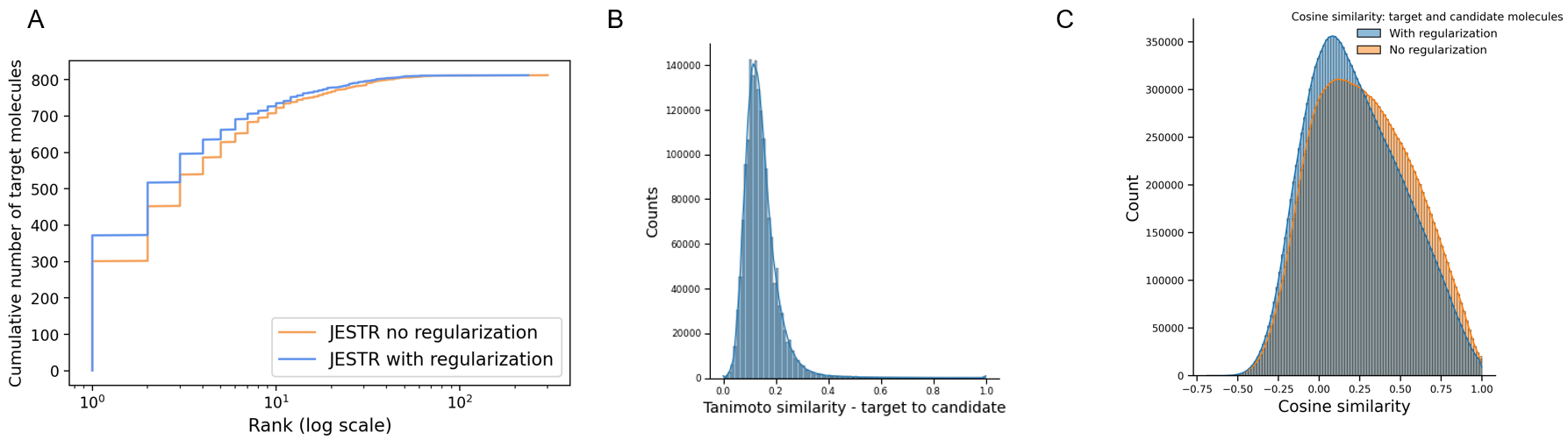}
  	\caption{Regularization analysis for JESTR for NPLIB1. A. Regularization improves rank@k by significantly placing more targets at rank 1. B. Distribution on Tanimoto similarities on the ECFP fingerprints between target and candidates in the training set. C. Distribution on cosine similarities, with and without regularization, of the target and candidates within the test set. 
   }
  	\label{fig:res_aug}
    \end{center}
  \end{figure*}
 \subsection{Ablation study - removing regularization}
To assess the value of regularization, the model was retrained without. Regularizing the training loss with  molecular candidates improves rank@1 by 11.4\%, 6.0\% and 37.1\% on the NPLIB1, NIST2020 and MoNA datasets, respectively. Improvements using regularization are evident at almost all ranks and all datasets (Table \ref{tab:results}B). For rank@20 on NPLIB1, the results drop by 0.3\% since the rank@20 result for NPLIB1 is  high even without regularization, where  even a small change in a few targets changes the result slightly. \fix{On the MassSpecGym dataset, regularization did not consistently boost model performance. Specifically, we observe a slight decrease in performance across all rank@k's when evaluating candidates by mass, and similarly, a modest reduction in rank@5 performance when using candidates by formula. We suspect that the challenging split within this dataset limits the generalizability benefits provided by regularization.}
 
To further demonstrate the value of regularization, we performed additional analysis on the NPLIB1 dataset. With regularization, the number of target molecules ranked @1 increases significantly, from 301 to 375, causing a ripple effect in improving other rank@k numbers (Figure \ref{fig:res_aug}A). Further, we examined the Tanimoto similarity between molecules in the training set and their candidates. 
We retrieved  $15.86$ million candidates from PubChem based on the chemical formulas of the target molecules in the training set. The majority of these candidates show low Tanimoto similarity with the target molecules (Figure \ref{fig:res_aug}B). Hence, our fine-tuning regularization strategy and sorting the candidates by their cosine similarity to the target effectively prioritizes regularization with the most similar candidates. We approximately utilized $7$ million candidates for regularization during the last 3\% of training epochs. 
Upon examining  the cosine similarity distributions on the embeddings of  candidate and target molecules (Figure \ref{fig:res_aug}C), we see that our regularization strategy reduces the average cosine similarity between targets and their candidates. Regularization is therefore  effective, enabling the model to discriminate between  a target molecule and its candidates. Similar analysis for the NIST2020 and MoNA datasets is shown in Figures S4 and S5 in the Supplementary Information.

\section{Conclusion}
JESTR offers a novel implicit annotation paradigm that avoids the explicit generation of spectra, fingerprints, or molecular structures. 
As molecules and spectra are views of the same object,  embedding these views in a joint embedding space using contrastive learning provides a  performance advantage.
\fix{Evaluation across diverse datasets, such as NPLIB1, NIST2020, MoNA, and MassSpecGym, demonstrates clear performance advantages over current explicit annotation techniques, highlighting the robustness of implicit models using the joint embedding approach. Moreover, investigation on the contrastive loss used by JESTR reveals that InfoNCE with temperature-scaled cosine similarity is advantageous for contrasting views and learning to distinguish matching pairs from non-matching pairs.}
% \fix{PLEASE FIX THIS @YAN ---   more meaningful summary really highlighting why the performance was not consistently higher.. or remove?? On NPLIB1, JESTR outperforms ESP by 76.12\%, MIST by 40.19\% and CMSSP by 6.59\% on ranks@[1-5].
% On NIST2020, JESTR outperforms ESP by 77.42\%, MIST by 43.57\%, and CMSSP by 1521.81\% on average on rank@[1-5]. On MoNA, JESTR outperforms ESP by 54.49\%, MIST by 1.66\% and CMSSP by 429.36\% on average on rank@[1-5]. On MassSpecGym, JESTR outperforms ESP by 26.91\%, MIST by 19.67\%, and CMSSP by 456.03\% on average on rank@[1-5]} 
Analysis of JESTR's performance on these datasets reveals that  dataset diversity, quality, and spectra-to-molecule ratios impact performance. Further, our results showed enhanced value in utilizing candidate molecules during training for regularization, improving performance by an average of \fix{ 5.72\%} for all datasets. \fix{Nonetheless, the minimal improvement in performance with regularization on the MassSpecGym benchmark suggests that the effectiveness of regularization may vary depending on data splits and generalization difficulty between the train and test set}. 
The overall JESTR results are promising and vouch for the potential of implicit annotation approaches. We expect to attain further improvements by  utilizing additional knowledge in the form of subformulae  annotation on spectral peaks and by utilizing enhanced molecular and spectral encoders.

%When trained and evaluated on the NPLIB1, JESTR offers  up to 93.2\% and 64.1\% improvement in rank@1 over the state-of-the-art mol-to-spec and spec-to-FP techniques, respectively. Our results also indicate the effectiveness of including candidate molecules to regularize the training, providing a up to 37.1\% improvement over a baseline that does not include regularization.

%The proposed approach relies on the accuracy of the chemical formula that can be identified from the spectra using techniques such as BUDDY \citep{xing2023buddy} or SIRIUS \citep{duhrkop2019sirius}. Without this knowledge, the candidate datasets would be retrieved based on molecular mass +/- error threshold.  While it is not possible to augment the training data with experimental spectra, augmentation with simulated spectra as was done in MIST \citep{goldman2023annotating} might have provided additional performance gain.

%\section*{Acknowledgments}
% \input{table_poster}

% \vspace*{-12pt}

\section*{Funding}
Research reported in this publication was supported by the National Intitute of
General Medical Sciences of the National Institutes of Health under award number  R35GM148219. The content is solely the responsibility of the authors and does not necessarily represent the official views of the  NIH.
%\vspace*{-12pt}

% Bibliography
\bibliographystyle{package/natbib}
\bibliography{references}

\end{document}

\title{\underline{\textbf{JESTR}}: \underline{\textbf{J}}oint \underline{\textbf{E}}mbedding \underline{\textbf{S}}pace \underline{\textbf{T}}echnique for \underline{\textbf{R}}anking Candidate Molecules for the Annotation of Untargeted Metabolomics Data

\textbf{Supplementary Information}}
\date{}

\author[1]{Apurva Kalia}
\author[1]{Yan Zhou Chen}
\author[2]{Dilip Krishnan}
\author[1,3]{Soha Hassoun}
\affil[1]{Department of Computer Science, Tufts University, Medford, MA 02155, USA}  
\affil[2]{Google Research}
\affil[3]{Department of Chemical and Biological Engineering, Tufts University, Medford, MA 02155, USA.}

\maketitle
\section{Batching strategy for contrastive learning and  regularization}
During contrastive learning, each batch of size $k$ contains $k$ target molecules. 
In each training epoch, the spectrum corresponding to any given target molecule is chosen sequentially from the list of paired spectra. The batch size $k$ was set to 32. The value of the temperature hyper-parameter $\tau$ was set to 0.05. For all experiments, contrastive learning was stopped at 1000 epochs, where it was observed that the contrastive loss  stopped reducing significantly. The batch size for the candidates, $k_{aug}$, was set to 32.

During regularization, for each training molecule in the batch, a set of most similar candidates for that molecule are used to calculate the regularization loss. This is done sequentially for every mini-batch in the epoch. A diagram outlining our batching strategy is shown in Figure S1.

\begin{figure}[b]
\center
	\includegraphics[width=\linewidth]{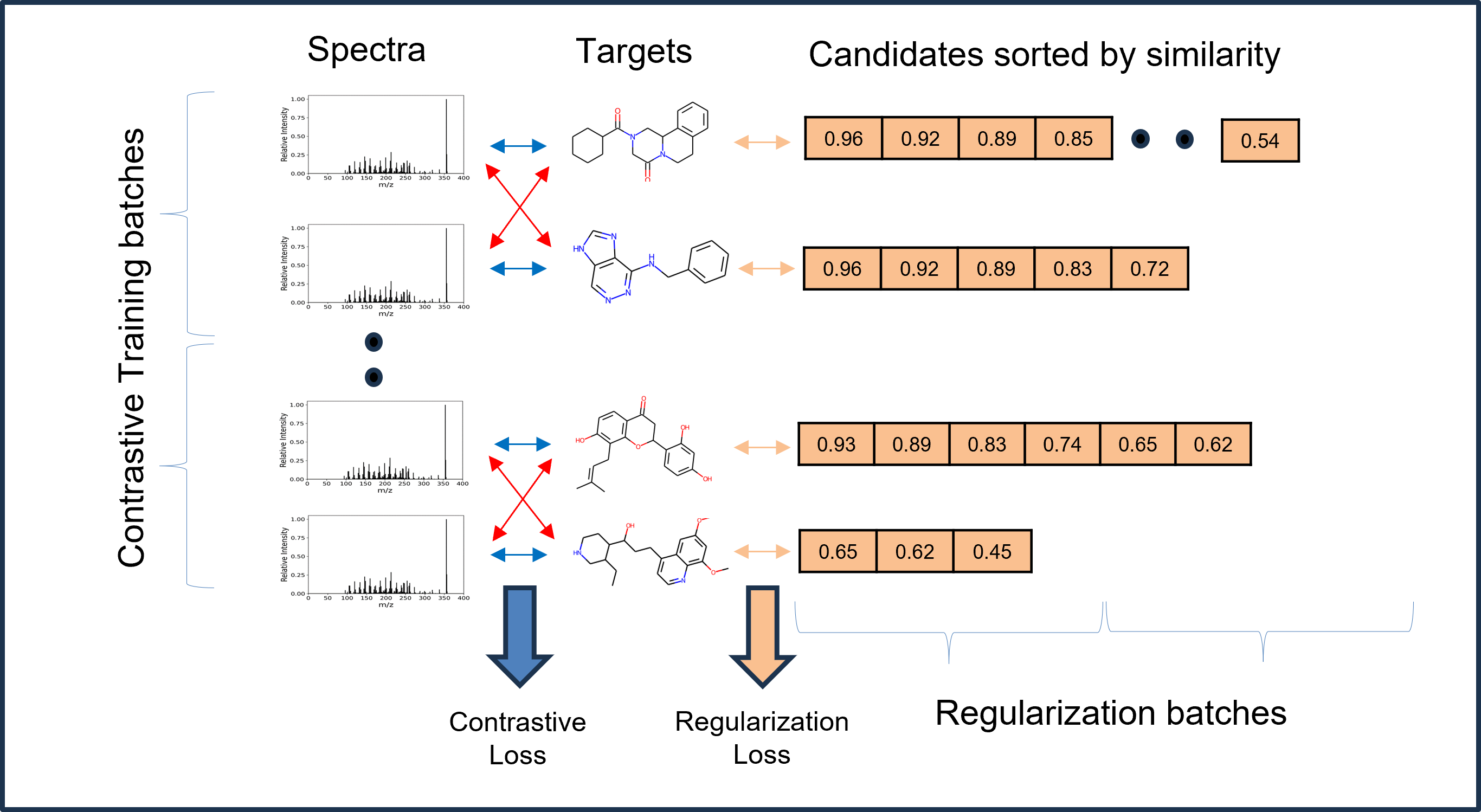}
	\caption{The contrastive loss is calculated over a batch of molecules at a time. For each molecule in the batch, candidates are selected from the candidate list sorted by similarity. If a particular molecule has a lesser number of candidates than the batch size, the candidates for that molecule will be repeated sequentially. The candidates thus selected are used to calculate the regularization loss. The curly braces in the vertical direction show the contrasting batching for two batches, while the curly braces in the horizontal direction show regularization batches for two batches}
	\label{fig:reg_batch}
\end{figure}

\section{Implementation details for JESTR}
JESTR was trained on NVIDIA A100 GPU with 40GB of graphics RAM and 256GB of CPU RAM. Adam \citep{kingma2014adam} was used as the optimizer. We used grid search over the ranges of the tuned parameters. The values of the parameters that achieved the best performance were selected and used to train and test the model for all datasets (Table \ref{tab:hp}).

\begin{table}[th]
\centering
\caption{Tuning of various hyperparameters for JESTR model using grid search.}
{\begin{tabular}{@{}lcc@{}}\toprule Hyperparameter &
Values searched & Final value\\\midrule
Contrastive Learning Rate & 5e-6, 5e-5, 5e-4 & 5e-4\\
Contrastive Batch Size & 32, 64, 128 & 32\\
Early Stopping Epochs & 20, 40, 80 & 80\\
Regularization loss weight & 0.1, 0.3, 0.5 & 0.1\\
\% epochs used for regularization & 3, 10, 20 & 3\\
\hline 
\end{tabular}\label{tab:hp}}{}
\end{table}

We also looked at the training and inference times of all the models (Table \ref{tab:runtime}). The runtimes  were measured on a Linux machine with 48 CPU cores with 196GB RAM and 6 nVidia A5000 GPUs with 24GB RAM. JESTR uses a GNN encoder for the molecules and an MLP encoder for the spectra. MIST uses MLP and transformers as its encoders for spectra, while ESP uses MLP and GNN as encoders for spectra and molecules, respectively. From a computational complexity of the model architecture, the three models are similar. The training time depends largely on the amount of training each model undergoes.

\begin{table}[th]
\centering
\caption{Training time and inference time for the 3 models on the NPLIB1 dataset. The runtimes were measured on the same machine. JESTR takes longer to train because the training is run for a larger number of epochs.} 
{\begin{tabular}{@{}lccc@{}}\toprule Model &
Training Time & Epochs & Inference Time\\\midrule
JESTR & 6 hours & 800 & 1.5 hours\\
MIST & 1.2 hours & 50 & 20 minutes\\
ESP & 1.3 hours & 100 & 1.2 hours\\
\hline 
\end{tabular}\label{tab:runtime}}{}
\end{table}

%The code and public NPLIB1 dataset are available on our GitHub \url{ https://github.com/HassounLab/JESTR1}

\section{JESTR ranks and target-candidate separation on NIST2020 and MoNA}
\begin{figure*}[tbh!]
\center
	\includegraphics[width=\linewidth]{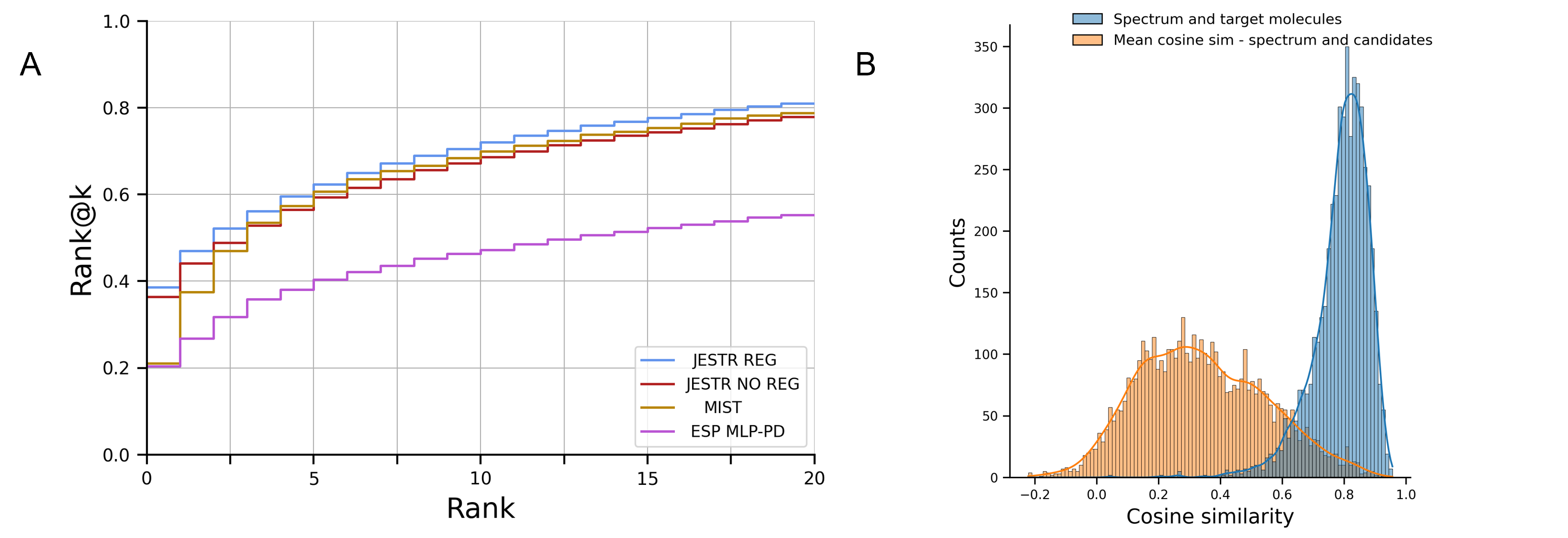} 
	\caption{Results. A. Rank@k results for JESTR on NIST2020, with and without regularization, MIST, and ESP MLP-PD. B. Distribution of cosine similarities of query spectra and target/candidate molecules in the NIST2020 test set with contrastive learning using JESTR. }
\label{fig:res_explicit_nist20}
\end{figure*}

\begin{figure*}[tbh!]
\center
	\includegraphics[width=\linewidth]{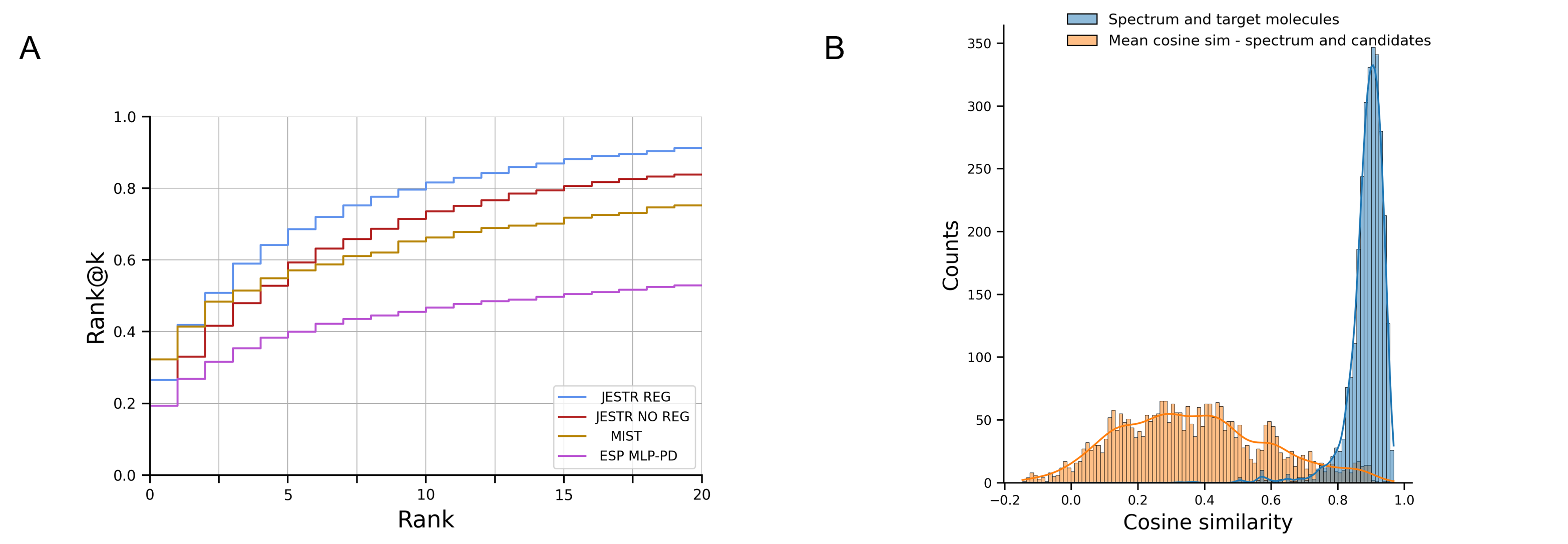} 
	\caption{Results. A. Rank@k results for JESTR on MoNA, with and without regularization, MIST, and ESP MLP-PD. B. Distribution of cosine similarities of query spectra and target/candidate molecules in the MoNA test set with contrastive learning using JESTR. }
\label{fig:res_explicit_mona}
\end{figure*}

We look at how JESTR separates target and candidate molecules for different datasets in Figures (\ref{fig:res_explicit_nist20} and \ref{fig:res_explicit_mona})

\section{Regularization analysis for NIST2020 and MoNA}
\begin{figure*}[btp!]
  	\includegraphics[width=\linewidth]{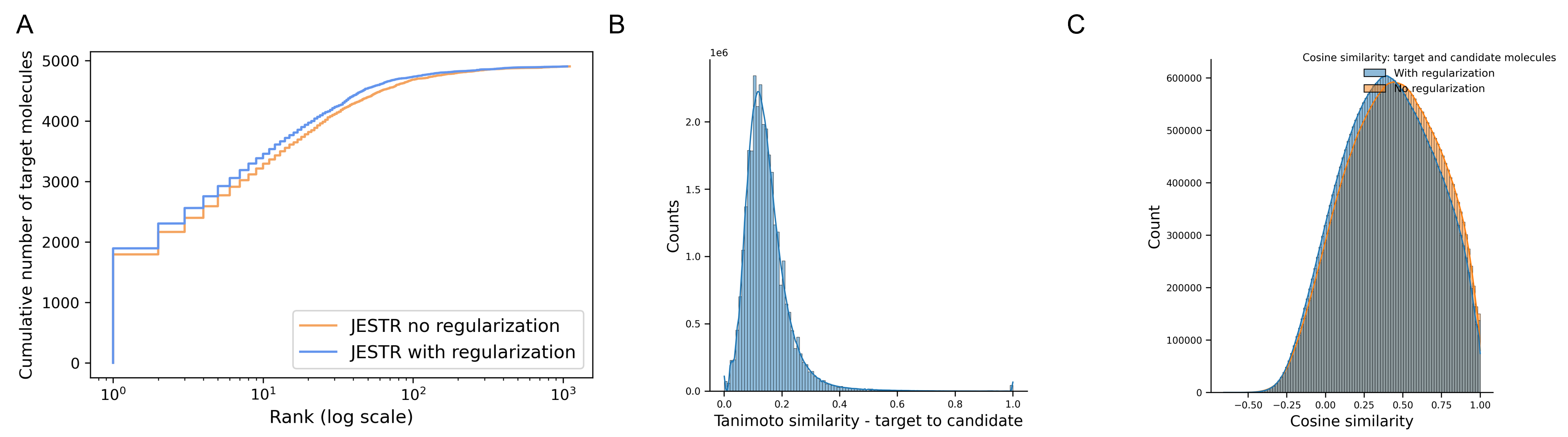}
  	\caption{Regularization analysis for JESTR for NIST2020. A. Regularization improves rank@k by significantly placing more targets at rank 1. B. Distribution on Tanimoto similarities on the ECFP fingerprints between target and candidates in the training set. C. Distribution on cosine similarities, with and without regularization, of the target and candidates within the test set. 
   }
  	\label{fig:res_aug_nist20}
  \end{figure*}

\begin{figure*}[t!]
  	\includegraphics[width=\linewidth]{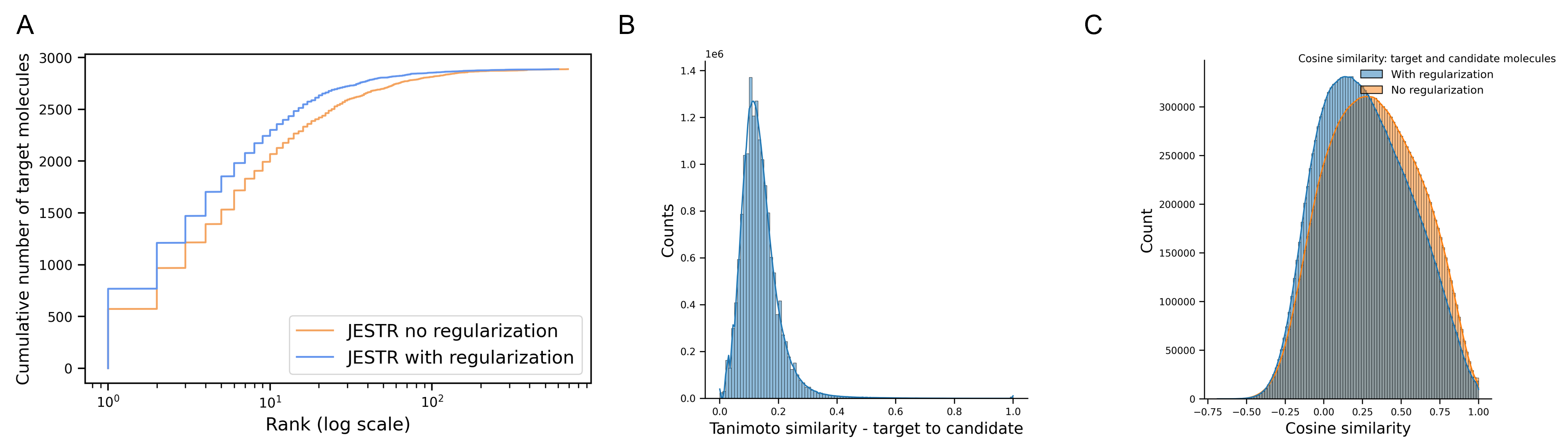}
  	\caption{Regularization analysis for JESTR for MoNA. A. Regularization improves rank@k by significantly placing more targets at rank 1. B. Distribution on Tanimoto similarities on the ECFP fingerprints between target and candidates in the training set. C. Distribution on cosine similarities, with and without regularization, of the target and candidates within the test set. 
   }
  	\label{fig:res_aug_mona}
  \end{figure*}

We study the impact of regularization on molecule ranking and on molecule embeddings for all datasets in Figures (\ref{fig:res_aug_nist20} and \ref{fig:res_aug_mona}).

\section{MCES Distributions between Train and Test Molecules}
\begin{figure*}[btp!]
    \includegraphics[width=\linewidth]{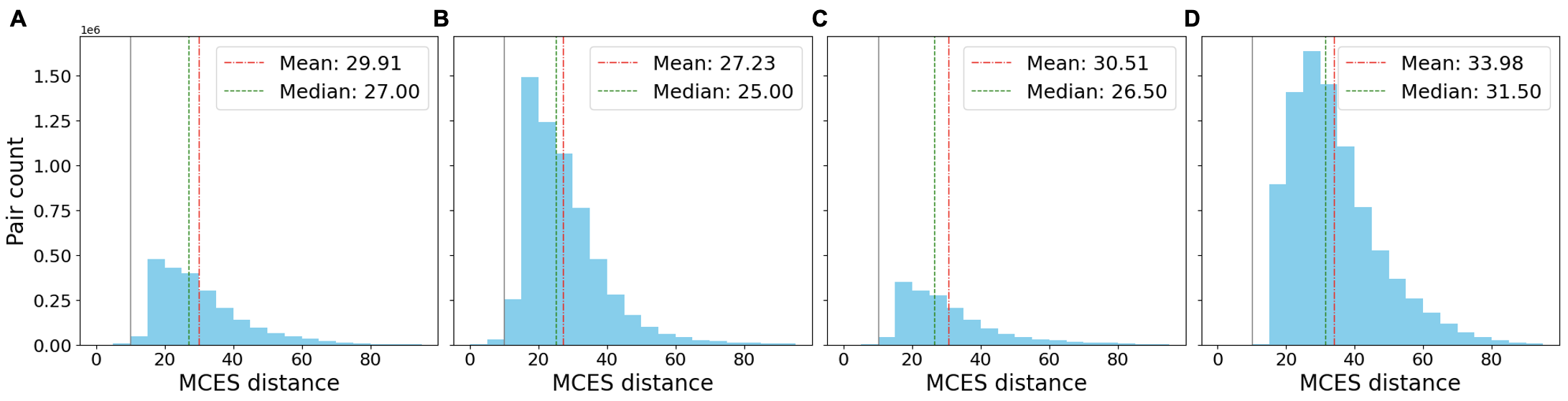}
    \caption{Distribution of MCES distances between molecules in SIRIUS's training set and test molecules from A)NPLIB1, B)NIST2020, C)MoNA, D)MassSpecGym. The solid gray line marks the MCES distance of 10.}
    \label{fig:mces_sirius}
\end{figure*}

\begin{figure*}[btp!]
    \includegraphics[width=\linewidth]{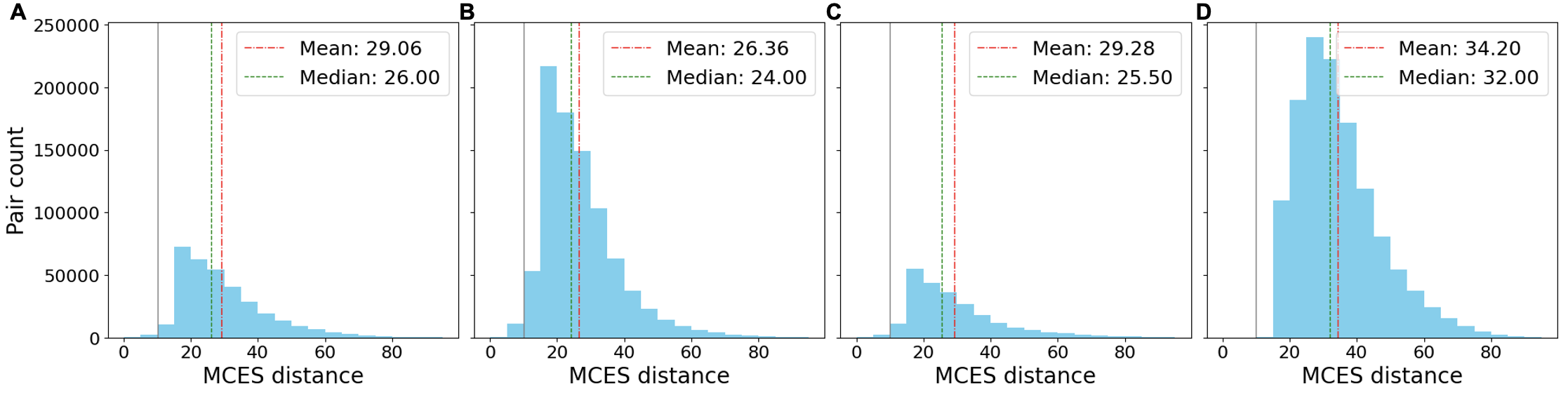}
    \caption{Distribution of MCES distances between molecules in CFM-ID's training set and test molecules from A)NPLIB1, B)NIST2020, C)MoNA, D)MassSpecGym. The solid gray line marks the MCES distance of 10.}
    \label{fig:mces_cfm}
\end{figure*}

\begin{figure*}[btp!]
    \centering
    \includegraphics[width=0.35\linewidth]{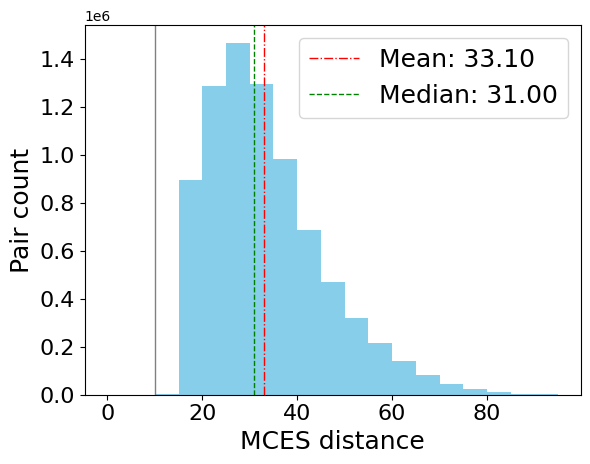}
    \caption{Distribution of MCES distances between train and test molecules in the MassSpecGym. The solid gray line marks the MCES distance of 10.}
    \label{fig:mces_msgym}
\end{figure*}

To fairly benchmark against SIRIUS and CFM-ID's pretrained models, we evaluate the distribution of the MCES distances between the pretrained molecules of SIRIUS and CFM-ID and the test molecules of NPLIB1, NIST2020, MoNA, and MassSpecGym. For each dataset, we randomly select 10\% of the train-test pairs and compute the MCES distance. We plot the distributions for SIRIUS in Figure \ref{fig:mces_sirius} and CFM-ID in Figure \ref{fig:mces_cfm}. The test set in MassSpecGym provides the most equitable setup, where the pretrained molecules are structurally different from the test molecules by an MCES distance of 10 or greater, just as the original MassSpecGym split (Figure \ref{fig:mces_msgym}).

\bibliographystyle{package/natbib}
\bibliography{supplementary}